\documentclass[twocolumn,showpacs,preprintnumbers,amsmath,amssymb]{revtex4}
\usepackage[dvips]{graphicx}

\begin{document}
\title{
Hole burning and higher order photon effects in attosecond light-atom
interaction}
\author{P. L. Price$^1$}
\altaffiliation{Now at Physics Department, University of Connecticut,
2152 Hillside Rd., Storrs Connecticut 06269-3046, USA}
\author{L. D. Noordam$^2$}
\author{H. B. van Linden van den Heuvell$^2$}
\author{F. Robicheaux$^1$}
\altaffiliation{Now at Department of Physics, Purdue University, West Lafayette,
Indiana 47907, USA}
\email{robichf@purdue.edu}
\affiliation{$^1$Department of Physics, Auburn University, AL
36849, USA}
\affiliation{$^2$Institute of Physics, University of Amsterdam,
PO Box 94485, 1090GL Amsterdam, The Netherlands}
\date{\today}

\begin{abstract}
We have performed calculations of attosecond laser-atom interactions
for laser intensities where interesting two and three photon effects
become relevant. In particular, we examine the case of ``hole burning"
in the initial orbital. Hole burning is present when the laser pulse
duration is shorter than the classical radial period because the electron
preferentially absorbs the photon near the nucleus. We also examine
how 3 photon Raman process can lead to a time delay in the outgoing electron
for the energy near one photon absorption. For excitation out of the
hydrogen $2s$ state, an intensity of $2.2\times 10^{16}$~W/cm$^2$ leads
to a 6 attosecond delay of the outgoing electron. We argue that this delay
is due to the hole burning in the initial state.
\end{abstract}

\pacs{78.47.J-, 32.80.Fb, 32.80.Rm}

\maketitle

\section{Introduction}

The past several decades have witnessed the increase in the speed of
lasers as tracked by the duration of a laser pulse. In recent years,
experimental groups have been able to decrease this duration into the
attosecond regime.\cite{PTB} With each decrease in duration of the laser
pulse, it is possible to probe quantum systems on shorter time scales.
For example, there have been many recent measurements and calculations
of the delay in the photoionization of electrons from atoms.\cite{SFK,KDG,GKA,NPF}
This example is interesting in that the delay is only of order 10~as but
can be measured using a streaking IR field {\it and} that
measurements and calculations differ in the expected delay; the
reason for the difference is not clarified. One of the results we
discuss below is that delays in electron ejection naturally occur
at higher laser intensities.

Although both the laser frequency and duration determine what types
of systems and phenomena are best investigated, there are types of
processes whose similarities extend from the attosecond to the
picosecond regime.
One of the processes that spans a broad regime is that a short
laser pulse can burn a hole in a wave function. The reason is that photoabsorption
is often highly position dependent. For example, stationary phase
considerations lead to the realization that photo-ionization
occurs near the nucleus within a region $r\sim \sqrt{(\ell +1)/\omega}$
with $\omega $ being the angular frequency of the laser.\cite{foot2}
Thus, a laser
pulse that is shorter than the radial period of the electron can
deplete the wave function in the neighborhood of the nucleus.\cite{NSD}

\begin{figure}
\resizebox{80mm}{!}{\includegraphics{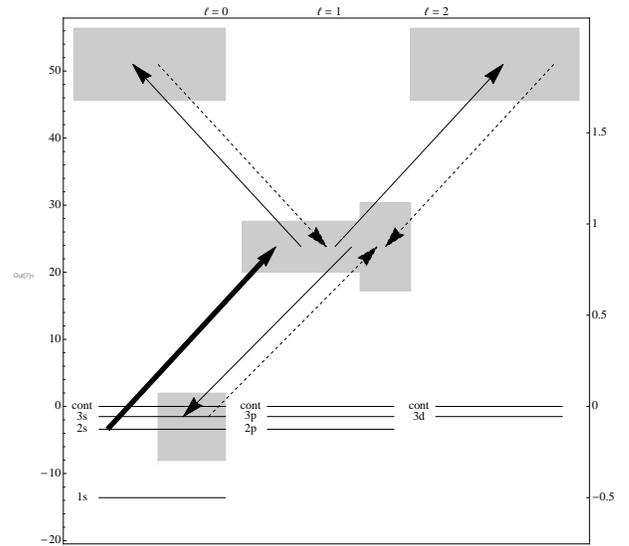}}
\caption{A schematic drawing of the energy levels and transitions
described in the paper.
}
\end{figure}

Figure~1 is a schematic of two photon processes. The gray bands are
meant to indicate the spread in energy for the processes that arise
from the short duration of the laser. The dark arrow is the one photon
excitation step. The narrow solid arrows are the two photon absorption
(for increasing energy) and the Raman process (for decreasing energy).
The Raman process back to $\ell =0$ gives the hole burning of the
initial wave function.
The dashed arrows indicate three photon process that can interfere
with the one photon absorption; the three photon processes can modify
both the energy distribution and the ejection time of the electron.

Hole burning was investigated in Rydberg states where
the radial period can be in the picosecond regime.\cite{HVN,DJ} For a state with
energy $E=-1/2\nu^2$, the radial period is given by
$T_{Ryd}=2\pi\nu^3$~a.u. which is
$T_{Ryd}= 1.52\times 10^{-4}\nu^3$~ps. Thus, states
with $\nu > 18.7$ have periods longer than a picosecond. One of the
manifestations of hole burning is Raman transitions to nearby states
with the same $\ell$. One way to think of this connection is that
the wave function with the hole, $\psi_h$, is projected onto the
eigenstates with the laser off. Since the $\psi_h$ mostly resembles
the initial state and has the same angular momentum,
the projection onto the initial state and states with
small change in $\nu$ and no change in $\ell$ will be emphasized.\cite{NSD}
The size of the hole in space increases with the duration of the laser
pulse. Thus, a short laser pulse will lead to a wider energy range
of final states. An equivalent statement in the frequency domain is
that short pulses
have a large bandwidth, supporting Raman transfer to neighbouring states.

In this paper, we examine the phenomenon of hole burning in the
regime of attosecond laser pulses. Since hole burning is at least a two
photon process, the intensity regime will be higher than what is
currently experimentally achieved but not so high that experiments
are unthinkable. While the basic phenomenon of hole burning should
be the same, there are differences due to the shorter time scale
and hence the lower states involved.
One difference is that for Rydberg states and picosecond
lasers, the hole burning leads to
population at both higher and lower energies (the principle quantum
number can both increase or decrease) while for attosecond lasers
the initial state is usually the ground state so population can
only transfer to higher energies. Another difference is that the
experiments on Rydberg state hole burning did not investigate when
the {\it ejected} electron is modified by the hole burning. In attosecond
physics, the time delay of the ejected electron is often one of
the more interesting parts of the measurement but also used to define
the `zero' time delay in pump-probe experiments. We give calculations
for this process and show that the hole burning leads to a delay
in the ejection of the electron. This effect arises from the interference
between one photon absorption and a three photon process (2 absorptions
and one emission).

We use atomic units unless SI units are explicitly given.

\section{Numerical Method}

We solved the time dependent Schr\"odinger equation by representing the
wave function on a grid of radial points and an angular momentum basis.
A more extensive discussion of the technique can be found in
Ref.~\cite{TR}.
In our calculations, we used linearly polarized light which leads to
a wave function that can be represented by the summation
\begin{equation}
\Psi (\vec{r})=\sum_\ell [R_\ell (r,t)]Y_{\ell m}(\Omega )
\end{equation}
where the radial functions, $R_\ell$, contain all of the useful
information about the wave function. The maximum $\ell$ in the sum
is determined by the duration, strength and frequency of the laser
and was chosen so that less than $10^{-9}$ of the population was
in $\ell_{max}$. For ease in solving the time
dependent Schr\"odinger equation, we consider the
Hamiltonian for this calculation as the sum of two terms:
\begin{equation}
H = H_1 + H_2 = H_{atom} + H_{field}
\end{equation}
where $H_1$ is the atomic Hamiltonian which contains the kinetic
energy operator and the spherical potential from the electron-ion
interaction and the $H_2$ is the laser-electron interaction.
For hydrogen, the electron-ion potential is simply $-1/r$. For the
other atoms presented in this paper, a model potential can be used.

The Hamiltonian for the laser
field was taken to be in the length gauge. We do not need to worry about
the increasing size of the laser potential with r because the laser
duration is so short the electron does not travel far before the laser
intensity returns to 0. We chose $H_2 = z E(t)$ where
$E(t)=-dA/dt$ with
\begin{equation}
A(t) = F(t) \frac{1}{\omega_0}\sin (\omega_0 t + \phi ).
\end{equation}
The $\phi$ is the carrier envelope phase
and $F(t)$ is a smooth function giving the envelope. We chose
this to be a Gaussian $F(t)=F_{max}\exp (-t^2/t_w^2)$ with
$F_{max}$ the maximum electric field when the carrier envelope
phase is zero.

We used a split operator method with a Crank-Nicolson approximation
to step the wave function by $\delta t$. The approximation is
\begin{equation}
\Psi(t+\delta t/2) = U_2(t,\delta t/2)U_1(\delta t)U_2(t,\delta t/2)\Psi (t-\delta t/2)
\end{equation}
where the
\begin{equation}
U_j(t,\delta t) = [1 - i H_j(t)\delta t/2][1+i H_j(t)\delta t/2]^{-1}
\end{equation}
gives $O(\delta t^3)$ accuracy for one time step. In the propagator,
we only need to perform the $U_2$ when $|t|<6t_w$ because $H_2$ is
approximately 0 outside of that range. Both the $H_1$ and
$H_2$ operators can be represented as tridiagonal matrices. For the
electric field, the operator $H_2$ only couples $\ell$ to $\ell\pm 1$.
We used two approximations for $H_1$, both of which gives a tridiagonal
representation, as a test of
convergence. In the simplest approximation, we used equally spaced
points in $r$ (i.e. $r_j = j\delta r$) and a three point difference for
the radial kinetic energy. The more complicated approximation used a
square root mesh and a Numerov approximation as in Ref.~\cite{FR1}.
In all calculations, we checked convergence with respect to the number of
radial points, the number of angular momenta, and the number of time
steps.

Before the laser turns on, the wave function is in an eigenstate. We find
the eigenstate by diagonalizing $H_1$ in a finite region $r<r_{max}$.
The $r_{max}$ is chosen to be large enough that the wave function can
not reach that distance while the laser is on. We also used the eigenstates of
$H_1$ to compute the energy distribution of the final state. The eigenstates
go to 0 at $r_{max}$ which leads to a discretized
continuum. The energy distribution
with angular momentum $\ell$ is approximated by
\begin{equation}\label{eqEdis}
P_\ell (\bar{E}) = \frac{|<a+1,\ell |\Psi >|^2 + |<a,\ell |\Psi >|^2}{2(E_{a+1,\ell}-E_{a,\ell})}
\end{equation}
where $E_{a,\ell}$ is the energy of the $a$-th eigenstate 
with angular momentum $\ell$,
$\bar{E}=(E_{a+1,\ell}+E_{a,\ell})/2$ is the average of the two energies,
and $<a,\ell |\Psi>$ is
the projection of that eigenstate on the wave function. We also use this
definition for negative energy states which
allows us to treat both positive and negative energy on the
same curve. This has the disadvantage that it is not clear what is the
population in an individual state, which is a measurable quantity.

\section{Two-photon processes}

In this section, we present results on two photon ionization and
on one photon absorption followed by one photon emission (and vice
versa). We investigate the latter process in more detail because
it is the mechanism that leads to ``hole burning" of the initial
orbital.\cite{HVN} It is important to remember that even a weak
laser pulse whose time width is shorter than the Rydberg period will
burn a hole in the wavefunction. The depth of the hole increases with
laser intensity and, for weak fields, the result of the hole burning
will scale with the square of the intensity. However, the states described
in this section are not populated by other mechanisms and, thus, can
be distinguished even for weak lasers.

\begin{figure}
\resizebox{80mm}{!}{\includegraphics{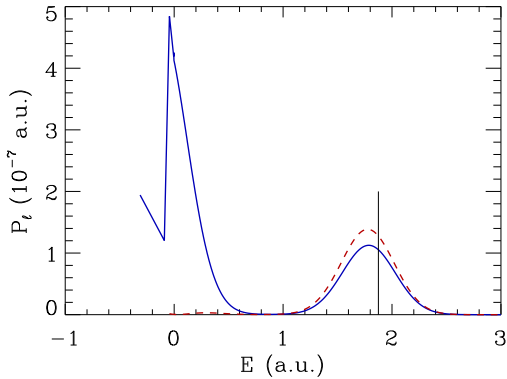}}
\caption{The energy distribution in atomic units of electrons for
$\ell = 0$ (solid line) and $\ell = 2$ (dashed line). 
In this calculation, we set
$\omega =1$, $F_{max}=0.05$~a.u. and $t_w=6.0$~a.u. In the
energy distribution formula, we set the population of the 2s state
to 0. The peak just below 2~a.u. is from two photon absorption; the
vertical line marks the position where a narrow band laser would have
the two photon ATI peak. The
Raman feature is from approximately -0.2 to 0.8~a.u. The $\ell =2$
Raman transition is very weak and is hardly visible on the graph.
The $\ell =0$ Raman structure is associated with hole burning in
the initial state.
}
\end{figure}

Figure~2 shows the energy distribution in atomic units for the 2-photon processes
when starting from the 2s state of H. For the presentation of the
energy distribution, we
set the population in the 2s to 0 when using Eq.~(\ref{eqEdis}) because
we want a plot of what states gained population; this choice leads to an
artificial dip in the Raman process. The solid line is the energy
distribution where the electron has $\ell =0$ and the dashed curve
is for $\ell =2$. In this calculation, we set
$\omega =1$ ($=27.2$~eV, $\lambda = 45.6$~nm),
$F_{max}=0.05$~a.u. and $t_w=6.0$~a.u.
The full width at half maximum (FWHM)
of the laser intensity is $t_{FWHM}=\sqrt{2\ln 2}t_w$ which corresponds
to $170\; as$ in this calculation. The electric field corresponds to
a maximum laser intensity of $8.8\times 10^{13}$~W/cm$^2$ 
which is in the perturbative regime. Doubling the intensity increases the
scale of Fig.~2 by a factor of 4. For this laser width, the state with
the most population is the $1s$ state; this is not obvious in Fig.~2
due to the large energy difference between the $1s$ and $2s$ states
which gives a large denominator in Eq.~(\ref{eqEdis}).

There are a few features worth noting in Fig.~2. The two photon absorption
peak is at nearly the same energy for the two angular momenta. However,
this energy, 1.78~a.u., is {\it not} at the expected energy for two
photon absorption which is $-1/8 + 2= 1.875$~a.u.; the expected energy
is marked by a vertical line in the graph. The peak is shifted
down in energy because the two photon absorption amplitude is a decreasing
function of energy; the product of the decreasing absorption amplitude and
the Gaussian centered at 1.875~a.u. gives the approximately 0.09~a.u.
shift. Using a longer laser pulse, leads to this feature becoming
narrower in energy with the peak shifting toward the expected value of
1.875~a.u. For example, the peak shifts to 1.83~a.u. when $t_w$ is
increased to 8~a.u. It is interesting that the 2-photon absorption 
has nearly equal population in $\ell = 0$ and 2 because the propensity
to increase $\ell$ when a photon is absorbed would lead to the expectation
that the $\ell =2$ would be substantially larger.

\begin{figure}
\resizebox{80mm}{!}{\includegraphics{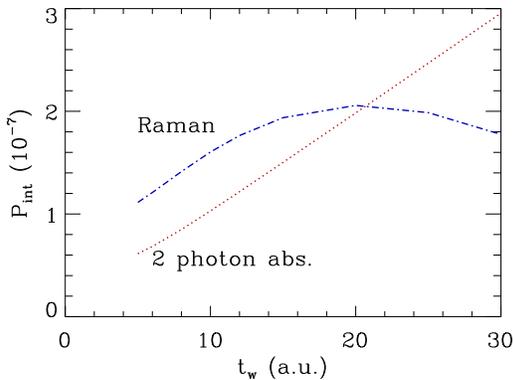}}
\caption{The total probability to be in the $\ell =0$ Raman structure
(dash-dot line) or the $\ell = 0$ two photon absorption peak (dotted
line). All
laser parameters are as in Fig.~2 except for the pulse width, $t_w$,
which is varied. The two photon absorption is approximately proportional
to the duration of the laser pulse. The probability for Raman transition
is not proportional to the pulse duration even for small $t_w$ and
begins to decrease when the bandwidth of the laser is too small to
reach the $3s$, $4s$, ... states. As a point of comparison,
$1/(E_{3s}-E_{2s})\simeq 14$~a.u. For pulses longer than $\tau_{2s}$
the bandwidth becomes insufficient to reach the $3s$ state in a two
photon Raman process.
}
\end{figure}

In the $\ell=0$ energy distribution, plotted in
Fig.~2, there is large population corresponding
to the two photon process
where one photon is absorbed and one photon is emitted. This process is
approximately 100$\times$ smaller for $\ell=2$. This disparity is somewhat
surprising considering that the two photon absorption probability is
comparable. The propensity rule would lead to the expectation that
$\ell =0$ would be larger than the $\ell =2$ but not by such a large
factor. We attribute the $\ell=0$ population
to hole burning of the initial orbital. The coherent superposition
of these Raman states with the initial state leads to a ``hole" in the initial
state at small $r$. Figure~3 shows the total probability ($\ell = 0$)
in the Raman transition and in the two photon absorption as the
pulse duration is increased keeping all other parameters fixed; the
data in this figure
comes from integrating the energy distribution up to $E=0.875$ for the
Raman transition and above $E=0.875$ for the two photon absorption.\cite{foot}
The
dependence of the two photon absorption on $t_w$ in Fig.~3 is the
simplest to understand. The probability is approximately proportional
to the duration because the ionization continues while the pulse is on.
The behavior of the Raman population is more interesting,
showing a slow increase followed by
a decrease. The Raman population starts decreasing when $t_w$ is comparable
to half the classical period; more specifically we find
$t_w\sim\sqrt{8\ln 2} n^3$. This condition is when the
energy FWHM of the laser field equals $1/n^3$, i.e. the energy spacing
of the atom. This fits with the interpretation of the Raman process
with $\Delta \ell = 0$ as equivalent to hole burning. When the laser
is on for a time comparable to or longer than the radial period for the
hole in the wave function, then there should not be any hole burning.

\begin{figure}
\resizebox{80mm}{!}{\includegraphics{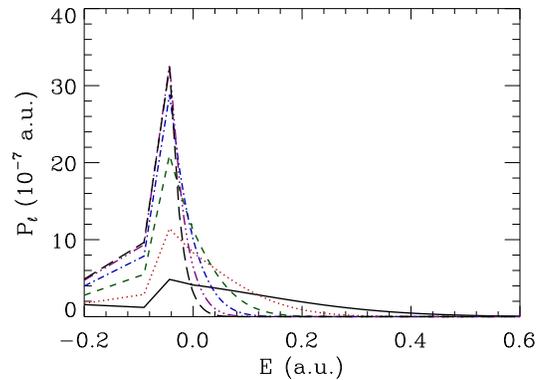}}
\caption{Same as Fig.~2 but only for $\ell = 0$ and
allowing $t_w$ to vary as in Fig.~3:
$t_w=6$~a.u. (solid line), 10~a.u. (dotted line), 15~a.u. (dashed line),
20~a.u. (dash-dot line), 25~a.u. (dash-dot-dot-dot line), and 30~a.u.
(long dash line). The longer duration gives a narrower energy distribution.
The probability to be in the states nearest $n=2$ increases with increasing
duration until the band width of the laser becomes too narrow to reach
those states.
}
\end{figure}

Figure~4 shows how the energy distribution for
$\ell =0$ Raman transition changes as the pulse duration is increased
keeping all other parameters fixed. One can see that the energy width
is decreasing while the peak is increasing for longer $t_w$. However,
the longest times have almost no increase in the peak while the width
continues to decrease. At even larger $t_w$, the Raman population at
all energies decreases until becoming approximately 0 when the bandwidth
of the laser is too narrow to allow any change in energy. As with Fig.~2
population in individual states are hard to visualize. The population
in the $1s$ state for each of the curves ($t_w=6$ through 30~a.u.) is
1.46E-7, 3.95E-8, 1.08E-9, 4.35E-12, 1.87E-15, and 1.87E-19 while the
population in the 3s state is 1.67E-8, 4.02E-8, 7.62E-8, 1.09E-7, 1.30E-7,
and 1.34E-7.

\begin{figure}
\resizebox{80mm}{!}{\includegraphics{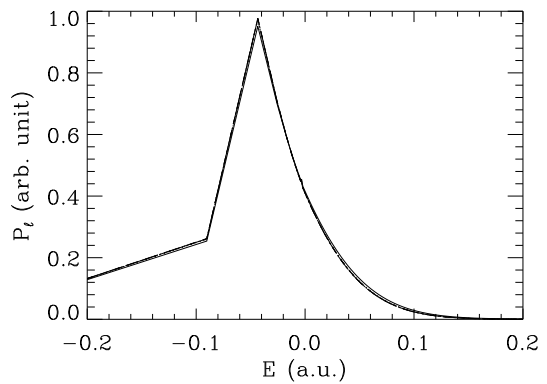}}
\caption{Same as Fig.~4 but fixing $t_w=18$~a.u. and allowing $\omega$
to vary. The integral of the curves have been fixed to be the same value. The
curves are $\omega = 0.33$~a.u. (solid line), 0.67~a.u. (dotted line), 1.0~a.u. (dashed line),
1.33~a.u. (dash-dot line), 1.67~a.u. (dash-dot-dot-dot line), and 2.0~a.u.
(long dash line). The lines are hardly distinguishable even though the
total Raman probability decreases by more than 4 orders of magnitude
from $\omega =0.33$ to 2.0~a.u. This shows the shape of the energy distribution
mainly depends on the laser duration and, hence, the shape of the hole does not
strongly depend on the laser frequency.
}
\end{figure}

Figure~5 shows how the energy distribution for $\ell = 0$ Raman transition
changes as the {\it frequency} changes keeping the pulse duration at
$t_w=18$~a.u. The shape of the energy distribution is nearly the same
for all frequencies shown even though there is a difference of more than four
orders of magnitude in the total population. This fits with the
interpretation of ``hole burning" since the shape of the hole will
mainly depend on the duration of the pulse and only weakly on the
laser frequency; the Raman transitions arise from the projection of
the hole in the initial wave function onto neighboring states.
We note that the central energy of the 2-photon absorption peak shifts
by $\sim 5/3$~a.u. for the same range of frequencies shown in Fig.~5.

We performed calculations for other single valence atoms and found
similar results. An exception is when the single photon transition
is at a Cooper minimum. The single photon ionization is strongly
suppressed in this situation while the two photon absorption is
enhanced compared to the Raman transition.

\section{One photon/Three photon ionization}

The ``hole burning" can affect the one photon absorption process by
changing the time when the electron is ionized and by changing the
energy distribution near the one photon absorption peak. Within a
perturbative picture, these effects result from the interference between
one photon absorption and a three photon process (absorb-emit-absorb).
The effect of ``hole burning" should be to delay the photon absorption
because the wave function is being depleted near the nucleus which is
where photoabsorption takes place; the photoabsorption is delayed for
strong laser fields because the electron probability near the nucleus
needs to refill which requires time.
Unlike the Raman process which is present even for weak laser
fields, this effect should only become apparent
at higher laser intensity reflecting a substantial change in the wave
function.

For this section, we fixed the duration, $t_w=6$~a.u., and
frequency, $\omega = 1$~a.u., of the laser pulse. In all calculations,
we started from the $2s$ state of hydrogen. The general trends
did not depend strongly on the type of atom or initial state.

\begin{figure}
\resizebox{80mm}{!}{\includegraphics{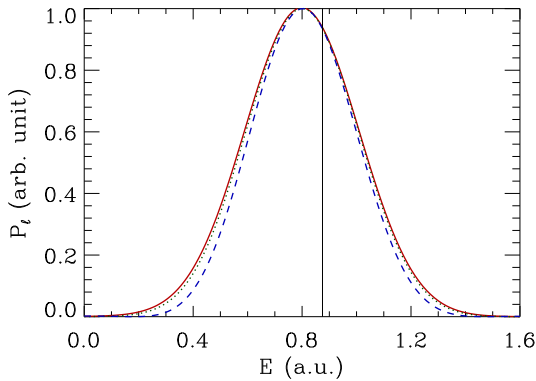}}
\caption{Same as Fig.~2 but for $\ell = 1$ and allowing $F_{max}$
to vary; the graph shows $F_{max}= 0.2$ (solid line), 0.4 (dotted line) and
0.8~a.u. (dashed line). The maximum value has been set to 1 for all
curves. The energy distribution for the highest laser intensity
is narrower and is slightly shifted to higher energies. The vertical
line marks the energy $E_{init}+\omega$ where a narrow band laser
would give the one-photon ionization.
}
\end{figure}

Figure~6 shows the energy distribution for $\ell=1$ near the one
photon absorption peak for $F_{max}=0.2$, 0.4 and 0.8~a.u. These
fields correspond to intensities of $1.4\times 10^{15}$,
$5.6\times 10^{15}$, and $2.2\times 10^{16}$~W/cm$^2$. The maxima
have been scaled to equal 1 for all $F_{max}$.
As with the two photon absorption peak in Fig.~2,
the position of the peak, $\simeq 0.8$~a.u., is slightly shifted to lower
energy compared to the expected value: $1 - 1/8=0.875$~a.u. which is
marked by a vertical line. The explanation
of the shift is the same: the dipole transition matrix element is a decreasing
function
of energy. While the smaller $F_{max}$ have nearly the same shape, the
energy distribution is narrower and shifted slightly higher for
the $F_{max}=0.8$~a.u. case. This effect must be due to the interference
between one photon absorption and a three photon process (either
emit-absorb-absorb or absorb-emit-absorb or absorb-absorb-emit).
By scaling all curves to have a peak of 1, we are hiding the fact that
the energy distribution increases with laser intensity. However,
we did not find that the increase was proportional to the laser intensity:
the ratio of scaling factors for the highest and lowest intensities
was 12 instead of 16 which is expected from perturbation theory.

As a point of comparison for the largest $F_{max}$, we compared
other processes to that shown in Fig.~6. The $p$-wave three photon
absorption is 240 times smaller.
The $f$-wave energy distribution is approximately
a 1000 times smaller for three photon absorption and
of 2000 times smaller for the $f$-wave peak near 0.8~a.u. These small
values might be surprising compared to the large nonperturbative
character for the ``one-photon" peak (e.g. the one-photon absorption
was approximately 25\%
too small at $F_{max}=0.8$~a.u.). However, it must be remembered that
the three photon processes can interfere with the one photon process which
can lead to a larger effect: the probability for an effect arising
from interference of one- and three-photon transitions
is $P=|A_1+A_3|^2 = |A_1|^2+(A_1A_3^*+A_1^*A_3) + |A_3|^2$. The interference
term is proportional to the square of the laser intensity while pure
three-photon processes are proportional to the cube of the laser intensity.
We note that the energy spread for the 3-photon absorption peak is
$\sim 50$\%
broader than the 1-photon peak; the extra broadness arises because the
higher order process depend on higher powers of the intensity and, thus,
they appear to be effectively shorter in duration. A simplistic argument
would suggest that the 1 photon peak would get broader at higher intensity
because of the addition of 3 photon character. The narrowing we
actually observe must be due to interference.

\begin{figure}
\resizebox{80mm}{!}{\includegraphics{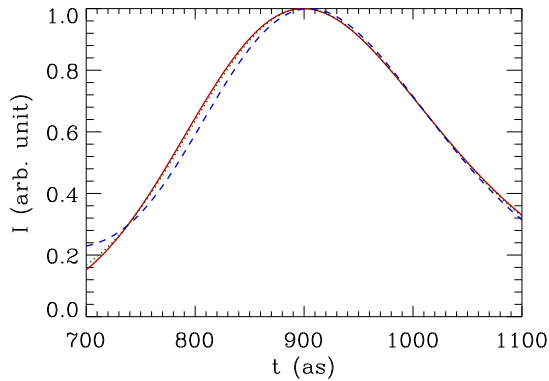}}
\caption{The electron
current at $r=50$~a.u. from all angular momenta; the line types have
the same meaning as Fig.~6.
All the times shown are after the laser field
is off. Note that the current at the highest intensity has slightly shifted
to later times which agrees with the expectation of changes due to
``hole burning".
}
\end{figure}

Figure~7 shows the time dependent current at $r=50$~a.u. with the peak
of the current scaled to be 1. A simple estimate of when the current
should peak is $t_{peak}\sim 50/\sqrt{2E}=39.5$~a.u.~$\simeq 953$~as. The
actual peak is somewhat earlier in time because the $-1/r$ potential
gives a higher radial speed to the electron; a classical electron
with energy 0.8~a.u. requires 907~as to reach 50~a.u. As in Fig.~6,
the smaller intensities give similar results while the highest intensity
gives a clear change. Part of the change is the current due to
two-photon absorption which leads to higher energy electrons and a peak
at early times; in a similar vein, there is a slow decay at the highest
intensity due to the Raman process which gives low energy
electrons that reach $r=50$~a.u. at later times.
A more interesting effect is that the one photon peak
is delayed by $\simeq 6$~as. Quantum mechanically this effect is due
to the interference between one- and three-photon paths. Physically,
the effect is consistent with the expectation of a time delay due to
hole burning in the wave function.

We note that the interference between one- and two-photon processes
in a single attosecond absorption has been discussed for both
energy and angular distributions. Reference~\cite{PTG} discussed the
effect on electron asymmetries, Ref.~\cite{PPS} discussed the effect
on momentum and energy distributions and Ref.~\cite{PSF} gave an
analysis based on perturbation theory. In our Fig.~2, one can see that
the one photon absorption peak which falls between the Raman and
2-photon absorption peak would overlap them in energy. This would lead
to left/right asymmetries in the electron angular distribution
as discussed in Ref.~\cite{PTG}.

\section{Conclusions}

We have performed calculations of attosecond absorption in one
electron systems to explore some of the possible two- and three-photon
effects. Our calculations demonstrate that one of the main two-photon effects is
a Raman type process that leads to a strong redistribution of population
with the same angular momentum as the initial state. We expect this
process to be common and is related to ``hole burning" of the initial
wave function. The calculations also show that the interference of one-
and three-photon paths can lead to changes substantially larger than
might be expected. We argued that the time delay in the one-photon
absorption with increasing laser intensity is also due to ``hole burning"
of the wave function.\cite{HVN} Because the absorption mainly occurs near the
nucleus, a hole there will decrease the ionization rate and lead to a
delay of the photoabsorption.

In this paper, we focused on the hole burning in a one electron system.
There can be other types of hole burning in two (or more) electron systems.
An example involving a short range perturber coupled to a Rydberg
series was investigated using the Ba $6snd$~$^{1,3}D_2$ states
perturbed by the $5d7d$~$^1D_2$ state as the example.\cite{VHN,DJ}
As an example requiring attosecond
lasers, the Be ground state is nominally $2s^2$ in the valence
shell. Calculations that include correlation have found that there is
a large admixture of $2p^2$ in the ground state wave function. Typically,
the photo-ionization is different from $2s$ and $2p$ orbitals. This leads
to the possibility of having a hole burned in the radial coordinate (as
described above) and a hole burned in the correlation. This will lead
to substantial Raman transition to the nominally $2p^2$ autoionizing
state. Thus, other types of wave packets can be initiated using hole
burning.

This work was supported by the Chemical Sciences, Geosciences, and
Biosciences Division of the Office of Basic Energy Sciencs,
U.S. Department of Energy and by NWO. We acknowledge support
from the EU Marie Curie ITN COHERENCE Network.


\begin{references}
\bibitem{foot2} This result is derived by setting the radial momentum
for angular momentum $\ell +1$ and energy
$E_{gnd}+\omega$ equal to the initial radial momentum.
\bibitem{foot} The way we grouped the Raman and 2-photon absorption is not accurate
for $t_w$ smaller than the data shown because the 2-photon absorption
begins to have probability below $E=0.875$ and the Raman transition
has probability above $E=0.875$.
\end{references}
\end{document}